\documentstyle[12pt,epsf]{article} 
\newcommand{\vkappa}{{\mbox{\boldmath$\kappa$}}}
\newcommand{\vq}{{\bf q}}
\newcommand{\vrho}{{\mbox{\boldmath$\varrho$}}}

\date{}
\begin{document}

\title{Processes with the $t$-channel singularity in the
physical region: finite beam sizes make cross sections
finite}

\author{K.~Melnikov\\{\em Institut f\"ur
Physik,Universit\"{a}t Mainz}\thanks{ D 55099 Germany, Mainz,
Johannes Gutenberg Universit\"{a}t, Institut f\"{u}r Physik,
THEP, Staudinger weg 7; e-mail: melnikov@dipmza.physik.uni-mainz.de}
\\and\\V.G.~Serbo\\{\em Institut f\"ur Theoretische Physik,
Universit\"at Leipzig }\thanks{Permanent address:
Novosibirsk State University, 630090, Novosibirsk, Russia;
e-mail: serbo@phys.nsu.nsk.su
}}

\maketitle

\begin{abstract}
It is  known  that some high-energy processes  have a $t$-channel
singularity in the physical region. In this paper  we show that
this singularity is regularized if one takes into account the
finite sizes of the colliding beams, i.e. the realistic situation
which takes place at high--energy colliders.  On this way we
obtain  the finite  cross section which is linear proportional to
the transverse sizes of the colliding beams.  As an application
of the above result  we study the production of the single $W$
boson at  $\mu^+\mu^-$ colliders in the reaction $\mu ^- \mu ^+
\to e \bar \nu _e W^+ $.  For the total energy $\sqrt {s}\approx
95$ GeV and reasonable parameters of the muon beams the
``beam--size '' dependent part of the cross section for this
reaction is of the order of the standard cross section of the
$\mu ^- \mu ^+ \to \mu^- \bar \nu _{\mu} W^+ $ process.
\end{abstract}

\vspace{1cm}

\begin{center}
MZ-TH/96-03
\end{center}

\newpage

\section {Introduction}

It is known since early 60's that some high--energy processes can
have a $t$-channel singularity in the physical region \cite
{Peirles}. This situation can occur if initial particles in a
given reaction are unstable and the masses of the final particles
are such that the real decay of the initial particles can take
place.

Recently in the paper \cite {Ginz} it was stressed that this
problem turned out to be a practical issue for the reaction $\mu
^- \mu ^+ \to e^- \bar \nu _e W^+$ . It is noted in \cite {Ginz}
that the standard calculation of this cross section leads to the
infinite result.  Indeed, if the invariant mass of the final
$e\bar \nu _e$ system is smaller than the muon mass $m$, the
square of the momentum transfer in the $t$-channel $q^2$ can be
both positive and negative depending on the scattering angles.
In the region of small $q^2$ defined by the inequalities
$-\Lambda < q^2 < \Lambda$ and $\Lambda \ll m^2$ the main
contribution to the discussed cross section is given by the
diagram with the exchange of the muonic neutrino in the
$t$-channel (see fig.1). The $t$-channel propagator brings the
factor $1/ q^2$ to the matrix element $M$ of this process

\begin{equation}
M \propto {1 \over q^2 + i \epsilon}
\label{1}
\end{equation}
which results in the power--like singularity in the standard
cross section
\begin{equation}
d\sigma \propto \int \limits _{-\Lambda} ^{\Lambda} |M|^2 \, dq^2
\propto B = \int \limits _{-\Lambda} ^{\Lambda} \, {dq^2 \over |
q^2 + i \epsilon |^2 } = \infty \, .
\label{2}
\end{equation}
It is, therefore, necessary to find physically reliable method to
produce definite prediction for the measurable number of events.

In the paper \cite {Ginz} (see also \cite {Peirles}) it was
suggested to regularize this divergence by taking into account
the instability of muons in the initial state.  If this is the
case, then the muon mass is the complex quantity with the
imaginary part proportional to the muon width $\Gamma $.  It is
possible then to solve the energy--momentum conservation
constraints with the conclusion that the square of the momentum
transfer in the $t$-channel acquires imaginary part proportional
to the muon width. This observation leads to the replacement
$$
q^2 \to q^2 -i\gamma,~~~~\gamma \sim m\Gamma .
$$
The divergent integral in (\ref {2}) is therefore regularized
\begin{equation}
B\,=\, \int \limits _{-\Lambda} ^{\Lambda} \, {dq^2 \over
| q^2 + i \epsilon |^2 } \, \to \,
\int \limits _{-\Lambda} ^{\Lambda} \, {dq^2 \over
| q^2 - i \gamma |^2 } \, \sim \, {1\over \gamma}
\sim  \, {1\over m \Gamma}.
\label{3}
\end{equation}
Consequently direct calculation of the scattering cross section
becomes possible.

We note in this respect that if such a regularization is
performed, it is possible to estimate the typical time necessary
for this reaction to occur.  It is not difficult to see that this
time is of the order of the moving muon life time, i.e. much more
larger than the time necessary for muon beams to cross each
other. Therefore for the realistic muon colliders the results 
of ref.\cite {Ginz} are not applicable and a new consideration is
necessary.

We have found that for the realistic muon colliders there exist 
much more important effect connected with the finite beam sizes. In
this paper we are going to show {\it that accounting for the
finite sizes of the colliding beams gives a finite cross sections
for the processes like}
\begin{equation}
\mu ^- \mu ^+  \to e^- \bar \nu _e X
\label{4}
\end{equation}
{\it  with the $t$-channel singularity in the physical region}.
This is the main result of our paper.  As an example, we
reconsider reaction  $\mu ^- \mu ^+ \to e \bar \nu _e W^+ $  and
show that the actual cross section of this process for the total
energy $\sqrt{s} \sim 100$ GeV is approximately $1$ fb for the
typical transverse beam sizes $a \sim 10^{-3}$ cm.

Let us mention here that the effect of the finite beam size at
the high--energy colliders is well studied both experimentally
and theoretically (for the review see ref. \cite {Obzor}). For
the first time this beam--size effect (BSE) was observed at the
VEPP-4 collider (Novosibirsk) in 1980--81 during the study of the
single bremsstrahlung in the electron--positron collisions
\cite{IYAF}.  This year the BSE was observed at HERA in the
reaction $ep \to ep\gamma $ \cite {HERA}.  In both cases the
number of observed photons was smaller than it was expected
according to the standard calculations. The decreased number of
photons was explained by the fact that impact parameters, which
gave essential contribution to the standard cross section of
these reactions, were larger by 2-3 orders of magnitude as
compared to the transverse beam sizes. From the theoretical point
of view, the BSE represents a remarkable  example of the
situation where traditional notion of the cross section and the
standard formulae for the number of events are not valid.

Let us briefly outline the principal points of our consideration.
The standard rules for calculations in high--energy physics
correspond to the collision of plane waves which represent
initial states of colliding particles. Instead of this we
consider collision of the wave packets which correspond to
the particle beams in the initial state. As a consequence the
standard rules are modified (see detailed discussion in the
section 3).  For instance, in the eq.(\ref {2}) the following
replacement takes place:
\begin{equation}
|M|^2 \, \to \, M_{fi}M^* _{fi'} \;.
\label{5}
\end{equation}
Here $M_{fi}$ and $ M _{fi'}$ are the matrix elements with
identical final but different initial states. The difference in
the momenta of the initial states is of the order of $1/a$ for
the beam of the transverse size $a$. Due to this fact the
singular propagators $1/q^2$ (which are the reason for the
pathological behaviour of the cross section) are now taken in the
different points of the momentum space for the amplitudes
$M_{fi}$ and $M_{fi'}$ :
$$
M _{fi} \propto  \frac {1}{q^2-\lambda + i \epsilon},~~~
M _{fi'} \propto \frac {1}{q^2+\lambda + i \epsilon},~~~
\lambda \sim \frac {m}{a}.
$$
Consequently the infinite factor $B$ is replaced by:
\begin{equation}
B = \int \limits _{-\Lambda} ^{\Lambda} \, {dq^2 \over
| q^2 + i \epsilon |^2 } \; \to \; \sim
\int \limits _{-\Lambda} ^{\Lambda} \, {dq^2 \over
(q^2 - \lambda +i \epsilon ) (q^2 + \lambda -i \epsilon ) }
\label{6}
\end{equation}
Below we  show that
\begin{equation}
B \sim {1\over \lambda} \sim {a\over m}
\label{7}
\end{equation}
and that the main contribution to this integral is given by the
pole of the neutrino propagator $q^2 +\lambda -i \epsilon =0$,
i.e. by the real neutrino.  This means \cite{Kotkin} that the
effect under discussion is connected with the decay $\mu^- \to
e^- \bar \nu_e \nu_\mu$ and subsequent scattering of the real
muonic neutrinos on the $\mu^+$ beam. As discussed in detail
below these neutrinos have a broad energy spectrum (see fig. 2)
\begin{equation}
dN_\nu = N_\nu \, f(x) \, dx
\label{8}
\end{equation}
($x$ is the fraction of the $\mu^-$ energy carried by neutrino).
In accordance with the eq.(\ref{7}), the total number of these
neutrinos (per one muon) is proportional to the typical
transverse beam size $a$:
\begin{equation}
N_\nu \, \sim \, {a\over c \tau}
\label{9}
\end{equation}
where $\tau = 1/ \Gamma$ is the muon life time, $c\tau = 660$ m.
As a consequence the  ``non--standard'' contribution to the
cross section is also proportional to the transverse beam size
\begin{equation}
\sigma_{\mbox{\scriptsize{non-stand}}} = N_\nu \int \, f(x) \sigma
_{\nu \mu \to W}(xs) \, dx \; \propto a.
\label{10}
\end{equation}

Let us mention here that in all the previous cases when the
effects of the finite beam sizes have been studied, the results
depend logarithmically on the beam size:
\begin{equation}
\sigma \propto \ln{a} \, .
\label{11}
\end{equation}
In contrast to this, in our case ( $t$-channel singularity in the
physical region ) the cross section (\ref{10}) is {\it linear}
proportional to the transverse sizes of the colliding beams.

The subsequent part of the paper is organized as follows:  in the
next section, as a realistic example of the processes with the
$t$-channel singularity in the physical region, we consider
reaction (\ref{4}) and present its cross section; in the 
section 3 we
prove the basic formula used in the section 2; section 4 is
devoted to the particular case --- the reaction $\mu^-\mu^+ \to e
\bar \nu _{e} W^+$; in the section 5 we present our conclusions;
Appendix is devoted to the detailed consideration of a model
process with a $t$-channel singularity in the physical region.

\section {General case }

In this section we show how the cross section of the reaction
$\mu ^- \mu ^+ \to e^- \bar \nu _e X$ can be calculated.  We
introduce the following notations:  $s=(p_1+p_2)^2=4E^2$ is the
square of the total energy in the center of mass frame, $m$ and
$\Gamma = 1/\tau$  are  muon mass and width respectively, $p_1^2
=p_2^2 =m^2$, $p_3$ is the 4-momentum of the final $e^- \bar \nu
_e$ system, $y=p_3^2/m^2 $, $q = p_1-p_3=(\omega, {\bf q})$ is
the momentum transfer in the $t$-channel and $x=qp_2 /p_1p_2
\approx \omega/E$.

From simple kinematics it follows that
\begin{equation}
q^2 = -{{\vq} _\bot {}^2\over 1-x} + t_0,~~~
t_0 \, =     \, { x(1-x-y)\over 1-x} \, m^2
\label{12}
\end{equation}
where  ${\vq} _\bot$ is the component of the momentum ${\vq}$
which is transverse to the momenta of the initial muons.
Note, that
\begin{equation}
t_0\, >\, 0 \;\;\; \mbox{for} \;\;\; y < 1-x
\label{13}
\end{equation}
and that for $q^2 =0$
\begin{equation}
|\vq _\bot| = q_\bot ^0 = m \sqrt{x(1-x-y)}.
\label {14}
\end{equation}

For $y< 1-x$ (or $t_0 > 0$) we write the cross section in the
form
\begin{equation}
d\sigma = d\sigma_{\mbox{\scriptsize{stand}}}+
d\sigma_{\mbox{\scriptsize{non-stand}}}
\label{15}
\end{equation}
where by definition $d\sigma_{\mbox{\scriptsize{stand}}}$
corresponds to the region $q^2 < -m^2$ and
$d\sigma_{\mbox{\scriptsize{non-stand}}}$ corresponds to the
region
\begin{equation}
-m^2 < q^2 < t_0 .
\label{16}
\end{equation}
We will show that the main contribution to $d\sigma
_{\mbox{\scriptsize{non-stand}}}$ is given by the region of a
very small values of $q^2$ inside the region (\ref{16}):
\begin{equation}
-\Lambda < q^2 < \Lambda, ~~~ m/a \ll \Lambda \ll m^2
\label{17}
\end{equation}
where $a$ is the typical transverse beam size.  In the region
(\ref{17}) the main contribution comes from the diagram with the
exchange of the muonic neutrino in the $t$-channel (fig.1).
Since for such $q^2$ the exchanged neutrino is almost real, the
corresponding matrix element can be considerably simplified.

First, the numerator of the neutrino propagator $\hat q$ can be
presented in the form
$$
\hat q \approx \sum _{\lambda} u_\lambda({\bf q})~
{\overline u}_\lambda({\bf q}).
$$
Here $u_\lambda({\bf q})$ is the Dirac spinor for the massless
fermion with momentum ${\bf q}$ and helicity $\lambda$.  Second,
in the region of interest the upper block of the diagram (see
fig.1) corresponds to the decay $ \mu ^- \to e \bar \nu _e \nu
_{\mu} $. In this decay only left-handed $\nu _\mu $ is produced.
Therefore in the above sum the only term with $\lambda = -1/2$ is
relevant.

As a result, in the region (\ref{17}) we present the matrix
element $M$ in the form
\begin{equation}
M=-M_{\mu\to e\bar \nu_e \nu_\mu}  ~{1\over q^2+i\epsilon}
~M_{\nu\mu \to X}
\label{18}
\end{equation}
where $M_{\mu\to e\bar \nu_e \nu_\mu}$ is the matrix element for
the $\mu^-$ decay and $M_{\nu \mu \to X}$ is the matrix element
for the $\nu _{\mu} \mu ^+ \to X$ process.  In both of these
subprocesses we take $q^2$ equal to zero.

\begin{figure}[htb]
\epsfxsize=8cm
\centerline{\epsffile{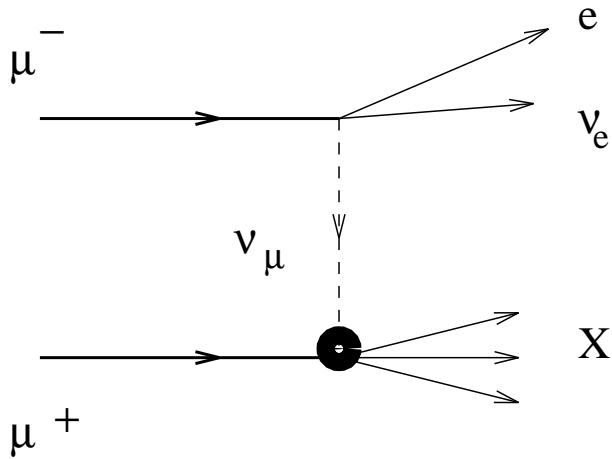}}
\caption[]{
The Feynman diagram for the reaction $\mu ^- \mu ^+ \to e \bar
\nu _e X $ which gives the leading contribution in the region of
small $|q^2|$.
}\end{figure}

Using this equation we express the non--standard contribution
through the muon decay width $\Gamma$ and the cross section
$\sigma_{\nu \mu \to X}$ of the $\nu_\mu \mu^+ \to X$ process as:
\begin{equation}
d\sigma_{\mbox{\scriptsize{non-stand}}} = \frac
{1}{\pi}~x~m~d\Gamma~\frac {dq^2}{|q^2|^2} ~d\sigma _{\nu \mu \to
X}.
\label{19}
\end{equation}
For  unpolarized muon beams we have
\begin{equation}
d\Gamma ={\Gamma \over \pi} (1-y)(1+2y)dxdy d\varphi
\label{20}
\end{equation}
where $\varphi$ is the azimuthal angle of the vector $\vq$.
Let us call the coefficient in front of $ d\sigma _{\nu \mu \to
X} $ in eq. (\ref {19}) as {\it the number of neutrinos}:
\begin{equation}
dN_{\nu}=\frac {1}{\pi}~x~m~d\Gamma ~\frac {dq^2}{|q^2|^2} =
{m\Gamma \over \pi^2}~x(1-y)(1+2y)dx dy d\varphi ~
{dq^2\over |q^2|^2}.
\label{21}
\end{equation}

As it is clear from the eq.(\ref {19}), the standard
calculation of the cross section turns out to be impossible due
to the power--like singularity, since the point $q^2=0$ is within
the physical region for $y < 1-x$.

The main result of our investigation  of the BSE in the above
process can be formulated as follows:

{\it Accounting for the BSE results in the following treatment of
the divergent integral in  the eq.(\ref {19})}:

\begin {equation}
B=\int \limits _{-\Lambda} ^{\Lambda} \,
\frac {dq^2}{|q^2|^2} \to \pi \frac {a}{q^0 _\bot} =
\pi\, {a\over m \sqrt{x(1-x-y)}}.
\label {22}
\end {equation}
The exact expression for the quantity $a$ will be given below
(see eqs.(\ref {42})--(\ref {45})).  We just mention here that it
is proportional to the transverse sizes of the colliding beams.
For the identical round Gaussian beams with the root-mean-square
radii
$$
\sigma _{ix}=\sigma _{iy}=\sigma _{\bot},~~i=1,2
$$
this quantity is equal to
\begin{equation}
a=\sqrt{\pi} \sigma _\bot.
\label{23}
\end{equation}

Hence, the quantity $B \sim a/m$ as it is noted in (\ref{7}). The
contribution (\ref{22}) comes from the region (\ref{17}). It is
not difficult to estimate that the contribution from the rest
part of the region (\ref{16}) is smaller. Indeed, its relative
value is of the order of
\begin {equation}
{1\over B}\int \limits _{-m^2} ^{-\Lambda} \,\frac{dq^2}{|q^2|^2}
\, + \, {1\over B}\int \limits _{\Lambda} ^{t_0}
\,\frac{dq^2}{|q^2|^2} \, \sim {m/a \over \Lambda} \, \ll \, 1 \,
.
\label {24}
\end {equation}

Using eqs. (\ref{22})--(\ref{24}) and integrating the number of
neutrinos (\ref{21}) over $y$ (in the region $0 < y < 1-x $) and
over $\varphi$, we arrive to the following spectrum of neutrinos:
\begin{equation}
\frac {dN_\nu (x)}{dx} = \frac {\pi a}{2c\tau} f(x),~~~
f(x)= \frac {24}{5\pi}
\sqrt{x(1-x)} \left(1+{22\over 9} x - {16\over 9} x^2 \right)
\label {25}
\end{equation}
where $\tau$ is the life time of the muon at rest, $c\tau =660$ m.
The function $f(x)$ is normalized to 1:
$$
\int \limits_{0}^{1} f(x)dx =1.
$$
The plot of this function is presented in the fig.2. The total
number of neutrinos is equal to
\begin{equation}
N_\nu = {\pi \over 2}\, {a \over c\tau}.
\label{26}
\end{equation}

After the spectrum of neutrinos is obtained, the non--standard
cross section for the reaction $\mu^-\mu^+ \to e\bar \nu _{e} X$
is given by the equation
\begin{equation}
d\sigma_{\mbox{\scriptsize{non-stand}}} = dN_\nu (x)\,d\sigma
_{\nu \mu \to X} (xs) = N_\nu \, f(x) \, dx \, d\sigma _{\nu \mu
\to X} (xs) \, .
\label{27}
\end{equation}

Subsequent integration over $x$ can be performed without
further difficulties. For the particular case $X=W^+$ such
calculation is performed in the section 4.

\section {Derivation of the basic formula}

Here we are going to prove  our result  presented in the previous
section (see eq.(\ref {22})).

To begin with, let us note that the standard notion of the cross
section is an approximation itself. As is well known, it
corresponds to the plane waves approximation for initial and
final particles. In the real experiments the particles are
confined to the beams of a relatively small size and it is the
collision of such beams that leads to the measurable number of
events.

In order to arrive to a more general formulae for the scattering
processes, it is necessary to describe the collisions of the wave
packets instead of the plane waves.  In view of the fact that the
movement of the particles inside a beam is quasiclassical, simple
and efficient technique for taking into account the beam--size
effects in the actual calculations has been developed \cite{KPS}
(for the review see \cite {Obzor}).

Below we present some results from the refs. \cite{KPS}, \cite
{Obzor} which are essential for our discussion. For simplicity,
we neglect the energy and angular spread of the particles in the
colliding beams.

Let us remind that in the standard approach the number of events
$N$ is the product of the cross section $\sigma $ and the
luminosity $L$:
\begin{equation}
dN=d\sigma~L,~~~ d\sigma \propto |M|^2, ~~~
L=v\int n_1({\bf r},t)~n_2({\bf r},t)d^3rdt
\label{28}
\end{equation}
where $v=|{\bf v}_1-{\bf v}_2| = 2$ for the head--on collision of
the ultra--relativistic beams.  The quantities $n_i({\bf r},t)$
are the particle densities of the beams.

The transformation from the plane waves to the wave packets
results in the following changes. The squared matrix element
$|M|^2$ with the initial state in the form of the plane waves with
the momenta ${\bf p_1}$ and ${\bf p_2}$ transforms to the product
of the matrix elements $M_{fi}$ and $M^*_{fi'}$ with the
different initial states:
\begin{equation}
d\sigma \propto |M|^2 ~~ \to ~~ d\sigma ({\mbox {\boldmath
$\kappa $}}) ~\propto ~ M_{fi}M_{fi'}^{*} \, .
\label{29}
\end{equation}
Here the initial state $|i \rangle $ is the direct product of the
plane waves with the momenta $ {\bf k}_1 = {\bf p}_1+\frac
{1}{2}{\mbox {\boldmath $\kappa $ }} $ and ${\bf k}_2 = {\bf p}_2
- \frac {1}{2}{\mbox {\boldmath $\kappa $ }}$, while the initial
state $|i' \rangle $ is the direct product of the plane waves
with the momenta ${\bf k}'_1 = {\bf p}_1- \frac {1}{2}{\mbox
{\boldmath $\kappa $ }}$ and ${\bf k}'_2 = {\bf p}_2+\frac
{1}{2}{\mbox {\boldmath $\kappa $ }}$.  Instead of the luminosity
$L$ the number of events starts to depend on the quantity
\begin{equation}
L({\mbox{\boldmath$\varrho$}} )= v\int n_1({\bf
r},t)~n_2({\bf r}+{\mbox {\boldmath $\varrho$}},t) \, d^3rdt
\label{30}
\end{equation}
through the following formula
\begin{equation}
dN = \int \frac {d^3\kappa d^3\varrho}{(2\pi)^3} \;
\mbox{e}^{ i{\mbox{\scriptsize\boldmath$\kappa$}} {\mbox
{\scriptsize\boldmath$\varrho$}}} \,
d\sigma
({\mbox {\boldmath$\kappa$}})\, L({\mbox {\boldmath$\varrho$}}).
\label {31}
\end{equation}

The characteristic values of $ \kappa $ are of the order of the
inverse beam sizes, i.e.
$$
\kappa \sim {1\over a}.
$$
Usually this quantity is much smaller than the typical scale for
the variation of the matrix element with respect to the initial
momenta. In this case we can put $\kappa =0$ in $d\sigma({\mbox
{\boldmath $\kappa $ }} )$ which results in the standard
expression for the number of events (\ref {28}).  Otherwise, one
should analyze the complete formulae which take into account the
effect of the finite beam sizes.

In view of the discussion given in the previous section, this is
indeed the situation which occurs in our case. Now we want to
show how the finite result for the number of events can be
obtained starting from the complete formula (\ref {31}).

Let us first define the ``observable cross section'' by the
relation (compare with the eq.(\ref{15}))
\begin{equation}
d\sigma =\frac {dN}{L} =
d\sigma_{\mbox{\scriptsize{stand}}}+ d\sigma
_{\mbox{\scriptsize{non-stand}}}
\label{32}
\end{equation}
where $L$ is the standard luminosity.  By writing the number of
events in such a way, we push the BSE to the quantity
$d\sigma _{\mbox{\scriptsize{non-stand}}}$.

The study of the matrix element of the discussed process (\ref
{18}) suggests that the only quantity sensitive to the small
variation of the initial momenta is the denominator of the
neutrino propagator for small values of $q^2$.  
Consequently, the transformations (\ref
{29})--(\ref{31}) reduces to the following modification:
\begin{equation}
\frac {1}{|q^2|^2} \to \frac {1}{t+i\epsilon}~\frac {1}{t'
-i\epsilon}
\label{33}
\end{equation}
where
\begin{equation}
q^2 = (p_1 -p_3)^2,~~~  t = (k_1-p_3)^2 , ~~~ t'= (k'_1-p_3)^2 .
\label{34}
\end{equation}
Let us expand $t$ and $t'$ up to the terms
linear in $\vkappa$. This gives
\begin{equation}
t=q^2- \lambda ,~~~ t'=q^2 + \lambda
\label{35}
\end{equation}
where
\begin{equation}
\lambda = \vkappa {\bf Q}, ~~~ {\bf Q} = -{\bf p}_3 + {E_3 \over
E_1}~ {\bf p}_1.
\label{36}
\end{equation}
In the center of mass frame of the colliding muons
\begin{equation}
{\bf Q} \approx -{\bf p}_{3\bot} = \vq _\bot = |\vq _\bot|\, {\bf
n}, ~~~ {\bf n} = (\cos{\varphi , \sin{\varphi, 0}}).
\label{37}
\end{equation}

As a result, the quantity $B$  in the eq. (\ref{22}) transforms to
(compare with the eq. (\ref{6}))
\begin{equation}
B = \int ~ \frac {d^3\kappa d^3\varrho}{(2\pi)^3} \;
\mbox{e}^{ i{\mbox{\scriptsize\boldmath$\kappa$}} {\mbox
{\scriptsize\boldmath$\varrho$}}} \;
\frac {L(\vrho)}{L}~
\int  \limits ^{\Lambda} _{-\Lambda} ~\frac {dq^2}
{(q^2- \lambda +i\epsilon)~(q^2 + \lambda -i\epsilon)}.
\label {38}
\end{equation}
It has already been mentioned that the above equation must be
understood in the sense that we isolate the part, which does not
possess the limit $\kappa \to 0$, to the quantity $B$. All 
other quantities are smooth in the above limit and hence do not
present any problem.

To perform the $q^2$-integration in (\ref{38}) we note \footnote
{The more detailed discussion of this integration can be found in
the Appendix.} that in the region (\ref{17})
\begin{equation}
\lambda = \vkappa {\bf n} \; \sqrt{(q^0_\bot )^2 - (1-x) q^2}
\approx  \lambda_0 = \vkappa {\bf n} ~ q^0_\bot .
\label{39}
\end{equation}
This approximation is justified because
$$
{|q^2| \over m^2} \stackrel{<}{\sim} {\Lambda \over m^2} \, \ll \,
1.
$$
Then we extend the region of integration over $q^2$ up to $\pm
\infty$. The error introduced by this procedure is of the
order of $(m/a) / \Lambda \, \ll \, 1$ (cf. eq.(\ref{24})).
Now the integral over $q^2$ along the real axis can
be replaced by the integral over the contour $C$ which goes around
the upper half plane of the complex variable $q^2$:
\begin{equation}
\int \limits _{-\infty}^{\infty} ~ {dq^2 \over D} =
\int \limits _{C} ~ {dq^2 \over D} = \, {\pi i \over -\lambda_0
+ i \epsilon}.
\label{40}
\end{equation}
Here
$$
D=(q^2- \lambda_0 +i\epsilon)~(q^2 + \lambda_0 -i\epsilon).
$$
The result is provided by the pole $q^2 + \lambda_0 -i\epsilon=0$ in
the upper half plane.

Further integrations are simply performed with the help of the
exponential representation
\begin{equation}
{i \over -\lambda_0 + i \epsilon} =
\int \limits _{0}^{\infty} \mbox{e}^{i\alpha
(-\lambda_0 + i \epsilon)  }\; d\alpha.
\label{41}
\end{equation}
Subsequent integrations over $\vkappa$ and $\vrho$  become
trivial and we obtain
$$
B= \pi \int \limits _{0}^{\infty} \;
{L(\alpha\, {\bf n}\, q_\bot ^0)\over L} \; d\alpha.
$$
After simple redefinition $\alpha q_\bot ^0= \varrho $
we get for the factor $B$:
\begin{equation}
B = \pi ~{a\over q_\bot ^0}, ~~~ a=\int \limits _{0}^{\infty}
\, \frac {L(\varrho {\bf n})}{L} \,d\varrho, ~~~
{\bf n} = {{\bf Q}\over |{\bf Q}|} \approx \frac {\vq
_\bot ^0}{q_\bot ^0}.
\label {42}
\end{equation}
This completes the proof of the eq. (\ref{22}).

At high energy colliders the distribution of particles inside the
beams can be often considered as Gaussian.  In this case
$L(\varrho {\bf n})$ equals:
\begin{equation}
L(\varrho {\bf n}) = L\exp \Big \{ -\varrho ^2 \Big ( \frac
{\cos^2\varphi}{2a_x^2}+\frac {\sin^2\varphi}{2a_y^2} \Big ) \Big
\}, ~~~ {\bf n}=(\cos\varphi,\sin\varphi, 0)
\label{43}
\end{equation}
where
$$
a_x^2=\sigma _{1x}^2+\sigma _{2x}^2, ~~~
a_y^2=\sigma _{1y}^2+\sigma _{2y}^2 .
$$
This results in the following expression for $a$:
\begin{equation}
a= \sqrt{\frac {\pi}{2}}~\frac { a_x~a_y}{\sqrt{a_y^2
\cos^2\varphi+ a_x ^2 \sin^2\varphi }}.
\label {44}
\end{equation}
For the round (but not identical) beams with the root-mean-square
radii
$$
\sigma _{1x}=\sigma _{1y}=\sigma _{1\bot},~~~
\sigma _{2x}=\sigma _{2y}=\sigma _{2\bot}
$$
we have
\begin{equation}
a = \sqrt{{\pi \over 2}}~\sqrt{\sigma^2 _{1\bot} + \sigma^2
_{2\bot}}.
\label{45}
\end{equation}
It is interesting to note that the quantity $a$ and the
non--standard cross section are determined by the size of the
largest beam. For the round and identical beams with $\sigma
_{1\bot} = \sigma _{2\bot} = \sigma _{\bot}$ the result of
(\ref{23}) can be obtained.

\section {The cross section of the
$\mu^-\mu ^+ \to  e \bar \nu _e W^+$ reaction}

Let us analyze a special example taking $X=W^+$.  The cross
section for the reaction $ \nu _{\mu} \mu ^+ \to W^+$ can be
written in the form
\begin{equation}
\sigma_{\nu  \mu  \to W} = 12\pi^2\frac {\Gamma(W\to \mu
\nu)}{M} \delta (xs-M^2)
\label{46}
\end{equation}
where $\Gamma(W\to \mu \bar \nu _\mu)=0.22$ GeV is the partial
$W$ decay width and $M= 80.2$ GeV is the $W$ boson mass.
Integration over the fraction of the neutrino energy $x$ in the
eq. (\ref{27}) becomes trivial and we obtain:
\begin{equation}
\sigma_{\mbox{\scriptsize{non-stand}}} (\mu^-\mu ^+ \to  e \bar
\nu _e W^+)= \frac {\pi a}{2c\tau}~\sigma _0~ x_0~f(x_0),~~~
x_0=\frac {M^2}{s},
\label{47}
\end{equation}
$$
\sigma_0=\frac {12\pi^2}{M^2}\frac {\Gamma(W\to \mu \nu)}{M}=20
~{\mbox{nb}}.
$$

For numerical estimates we take $a=\sqrt{\pi}\sigma_\bot$ (which
corresponds to the case of the round identical Gaussian beams)
with (see ref.\cite {Palm})
$$
\sigma _\bot=10^{-3}~ \mbox {cm}.
$$
This non--standard cross section reaches the maximum of $0.76$ fb
for $\sqrt{s}=93$ GeV.  For larger energies this cross section
decreases as $s^{-3/2}$. It is interesting to note that the
modest value of the non--standard cross section $0.76$ fb is the
result of the product of the very small number of neutrinos
$$
N_\nu ={\pi \sqrt{\pi}\over 2} {\sigma_\bot \over c\tau} = 4.2
\cdot 10^{-8}
$$
and the huge value of the cross section for the $\nu_\mu
\mu^+ \to W^+$ transition averaged over neutrino spectrum
$$
\langle \sigma_{\nu \mu \to W} \rangle = \sigma_0 \, x_0\, f(x_0)
= 1.8 \cdot 10^7 \; \mbox{fb},~~~~x_0=0.74.
$$

First, let us compare this non--standard piece with the standard
contribution to the same cross section.  We  remind that by
``standard'' contribution we mean the cross section of the same
reaction calculated by the standard rules excluding the region of
the final phase space where $q^2~>~-m^2$. This contribution was
calculated \cite {Private1} with the help of the CompHEP package
\cite {Boos}.  The comparison of both contributions is shown in
the fig.3.  It is seen that the non--standard contribution
dominates up to the energies $\sqrt {s} \approx 105$ GeV.

Second, we compare our non--standard cross section with the cross
section for the single $W$ boson production  in the reaction
$\mu^-\mu^+ \to \mu ^- \bar \nu _{\mu} W^+$.  The latter is
completely standard process since it has no $t$-channel
singularity.  The reasonable estimate of its cross section can be
quickly obtained with the help of the equivalent photon
approximation.  It gives the cross section $\approx 1$ fb at
$\sqrt {s} \approx 95 $ GeV (where it  almost coincides with our
non-standard cross section). At higher energies the process
$\mu^-\mu^+ \to \mu ^- \bar \nu _{\mu} W^+$ dominates as compared
with the discussed process $\mu^-\mu ^+ \to  e \bar \nu _e W^+$.

\section{Conclusions}

We have considered the processes with the $t$-channel singularity
in the physical region. As the result of our study the following
statement is proved:

{\it Accounting for the finite beam sizes eliminates the
singularity and makes it possible to obtain the finite result for
the observable cross section.}

In contrast to all previously known cases, the beam--size effect
for the processes with the $t$-channel singularity in the
physical region turns out to be {\it linear} proportional to the
transverse size of the colliding beams.

The effect appears to be connected with the  real decay of the
unstable particle in the initial state followed by the 
scattering of its decay products on the opposite beam
\cite{Kotkin}.  Evidently, the ``interaction region '' for such a
scenario is only bounded if the finite sizes of the colliding
beams are taken into account. In the traditional approach to the
scattering process the initial states are the plane waves, which
are not restricted to any finite volume. As a price,  infinite
cross section due to the $t$-channel singularity is obtained.

The results presented in this paper correspond to the situation
when the parameter $a/(c\tau) \sim N_\nu \ll 1$ or when the
unstable particle life time is much larger than the interaction
time ( for muon colliders this approximation is clearly perfect).
If this parameter increases, stationary
approximation becomes wrong at some point.  In this case the
effects of the initial particle instability and the beam size
effects should be considered simultaneously.

As a practical example, we consider the process of the single $W$
boson production at $\mu^-\mu^+$ colliders. We demonstrate that
the standard cross section for a single $W$ boson production and
the non--standard one are of the same order of magnitude in the
energy region slightly above threshold of this reaction.  For
higher energies the relative size of the non--standard
contribution rapidly decreases.

We also note that precisely in the same way as above the processes
like $\mu^-\mu ^+ \to  e \bar \nu _{\mu} W^+$ and
$\mu^-\mu ^+ \to  \nu _{\mu} \bar \nu _e W^+$ can be considered.
In this processes the $t$-channel singularity will be due to the
electron neutrino and electron propagators respectively. In virtue of the
discussion given above these processes will correspond to the collision of 
the real electron neutrino or electron produced in the $\mu ^-$ decay with
the opposite $\mu ^+$ beam.

\section{Acknowledgments}
K.M. is grateful to the Graduiertenkolleg ``Teilchenphysik'',
Universit\"at Mainz for support and to E.~Sherman for a number of
fruitful discussions.  V.G.S. acknowledges support of the
S\"achsisches Staatsministerium f\"ur Wissenschaft und Kunst,
of the Russian Fund of Fundamental Research and of INTAS 93-1180.
We thank V.S.~Fadin, I.F.~Ginzburg and G.L.~Kotkin for valuable
discussions.  We are grateful to E.E.~Boos and A.E.~Pukhov for
providing us with the results of the CompHEP calculations.

\vspace{0.1cm}
{\Large \bf Appendix: The model example}

\vspace{0.5cm}
In this appendix a simple example 
of the
reaction $2\to 2$ with the $t$-channel singularity in the
physical region is
considered in detail. The same notations as in the main text of the paper 
are used.
We work within the frame of the scalar field theory with the
interaction Lagrangian
$$
{\cal L }_I (x) =g\, A(x)B(x)C(x).
$$
The fields $A$, $B$ and $C$ represent the particles with the
masses $M$, $m$ and zero respectively. For definiteness  we
assume $M > m$.  To avoid confusion, note also that the coupling
$g$ has a dimension of mass. The particle $A$ is unstable due to the
$A \to B + C$ process, its  decay width is of the order
\begin{equation}
\Gamma \sim {g^2\over 16 \pi M}.
\label{a1}
\end{equation}

Let us consider the scattering of the particles $A$ and $B$ :
\begin{equation}
A(p_1)+B(p_2) \to B(p_3)+ A(p_4).
\end{equation}
The matrix element of this process corresponds to the Feynman
diagrams with the $C$-particle exchange in the $t$- and
$s$-channels. For high energies, $s \gg M^2,~m^2$, the dominant
contribution comes from the diagram with the  $C$-particle
exchange in the $t$-channel:
$$
M= -{g^2\over q^2 +i\epsilon}.
$$

Simple kinematical relations provide the following limits for
the square of the momentum transfer $q^2$ in the $t$-channel:
\begin{equation}
-s \le q^2 \le t_0=\frac {(M^2-m^2)^2}{s}.
\end{equation}
Hence, the value of $q^2$ goes through zero.
As was already indicated in the paper, in this case the cross
section can not be calculated by means of the standard methods.
It is easy to see, that $q^2=0$ for the value of the transverse
momentum
\begin{equation}
|{\bf q}_\bot| = q^0_\bot = \sqrt{t_0} = {M^2-m^2\over \sqrt{s}}.
\label{a2}
\end{equation}

To model the real situation discussed above, the following
conditions have to be fulfilled:

{\it (i)} The unstable particle life time must be much larger
than the interaction time, i.e. (compare eqs. (\ref{3}) and
(\ref{22}))
\begin{equation}
M\Gamma \ll {\sqrt{t_0}\over a}.
\label{a3}
\end{equation}

{\it (ii)} The transverse beam size $a$ must be large in comparison
with the typical  inverse
transverse momentum (\ref{a2}), i.e.
\begin{equation}
a \gg {1\over \sqrt{t_0}}.
\label{a4}
\end{equation}

Taking into account (\ref{a1}) and introducing  small
parameter $\delta$ we express both of these constrains in the
form
\begin{equation}
{g^2 \over 16 \pi t_0}~ \ll ~ \delta = {1\over a \sqrt{t_0}}~ \ll
~ 1.
\label{a5}
\end{equation}

We now apply the same technique, as in the main text of the
paper. We start from the  complete description of the scattering
process in terms of the colliding beams (see eqs.(\ref
{29})--(\ref{32})) and calculate the quantity $\sigma
(\vkappa)$. For this aim we expand $t$-channel propagators up to
the terms of the order $\vkappa $, exactly as it has been done in
the main text.  Then the following formulae are valid:
\begin{equation}
M_{fi} = -{g^2 \over t +i\epsilon}, ~~~~
M_{fi'} = -{g^2 \over t' +i\epsilon}
\label{a6}
\end{equation}
where the quantities $t$ and $t'$ are given
in (\ref{34})--(\ref{37}).
To calculate
$$
{d\sigma(\vkappa)\over d\varphi} = \int\limits ^{t_0}_{-s} \;
{M_{fi}M^*_{fi'} \over 32 \pi^2 s^2}\; dq^2
$$
we have to calculate the integral
$$
J= \int\limits ^{t_0}_{-s} \; {dq^2 \over D(q^2)}, ~~~
D(q^2) =(t +i\epsilon)(t' -i\epsilon) =(q^2-\vkappa \vq_\bot
+i\epsilon ) (q^2+\vkappa \vq_\bot -i\epsilon ).
$$
Here the quantity
\begin{equation}
\vkappa \vq_\bot =\vkappa {\bf n} \sqrt{t_0 - q^2}
\label{a7}
\end{equation}
depends both on $q^2$ and $\varphi$.
To evaluate $J$ in the limit of the small momenta
$\vkappa$, we rewrite it in the following way:
\begin{equation}
J= \int \limits_{C}^{} \,{dq^2 \over D(q^2)}
-\int \limits_{-\infty}^{-s} \,{dq^2 \over D(q^2)}-
\int \limits_{t_0}^{\infty} \,{dq^2 \over D(q^2)}
\label{a8}
\end{equation}
where the contour $C$  goes around the upper half plane\footnote {Note, 
that due to the eq.(\ref {a7}) the last integral in the eq.(\ref {a8})
also requires some definition. However for our purposes the dependence
of this integral on $\vkappa$ is not important. Hence the last integral
in the eq.(\ref {a8}) should be understood as the value of this integral 
for $\vkappa =0$.}.

The integral over the contour $C$ can be
evaluated by using Cauchy's theorem.  The result is provided by
the pole in the upper half plane
\begin{equation}
\int \limits_{C}^{} \,{dq^2 \over D(q^2)} = {2\pi i \over R}
\label{a9}
\end{equation}
where
\begin{equation}
R= {dD(q^2) \over dq^2} ~~~ \mbox{at}~~~
q^2+\vkappa \vq_\bot -i\epsilon = 0.
\label{a10}
\end{equation}
Taking into account the terms of the order  $\delta$ and
neglecting the terms of the order  $\delta^2$, we obtain
\begin{equation}
R=2(-\vkappa {\bf n} \sqrt{t_0} +i\epsilon).
\label{a11}
\end{equation}
As for the integrals both from $t_0$ to the $\infty $ and from the
$-\infty$ to $-s$, we note that the dependence of the integrand
on $\vkappa$ can be neglected and consequently
\begin{equation}
\int \limits_{-\infty}^{-s} \,{dq^2 \over D(q^2)}+
\int \limits_{t_0}^{\infty} \,{dq^2 \over D(q^2)} \approx
{1\over t_0}.
\label{a12}
\end{equation}

As a result, we obtain
\begin{equation}
{d\sigma(\vkappa)\over d\varphi} = {g^4 \over 32 \pi^2
s^2}~ \left(  {\pi i \over -\vkappa {\bf n} \sqrt{t_0}
+i\epsilon } -{1\over t_0} \right).
\label{a13}
\end{equation}
Using representation (\ref{41}) we get after trivial
integration over $\vkappa$ and $\vrho$
\begin{equation}
{d\sigma (AB \to BA)\over d\varphi} = {g^4 \over 32 \pi^2
s^2}~ \left(  {\pi a \over \sqrt{t_0} } - {1\over t_0} \right)
\label{a14}
\end{equation}
where the quantity $a$ is defined in (\ref{42}). Generally
speaking, $a$ can depend on the angle $\varphi$ (see (\ref{44})).
Note that the "beam--size" dependent part dominates in the cross
section (\ref{a14}) due to the inequality (\ref{a4}).

\begin{figure}[htb]
\epsfxsize=14cm
\centerline{\epsffile{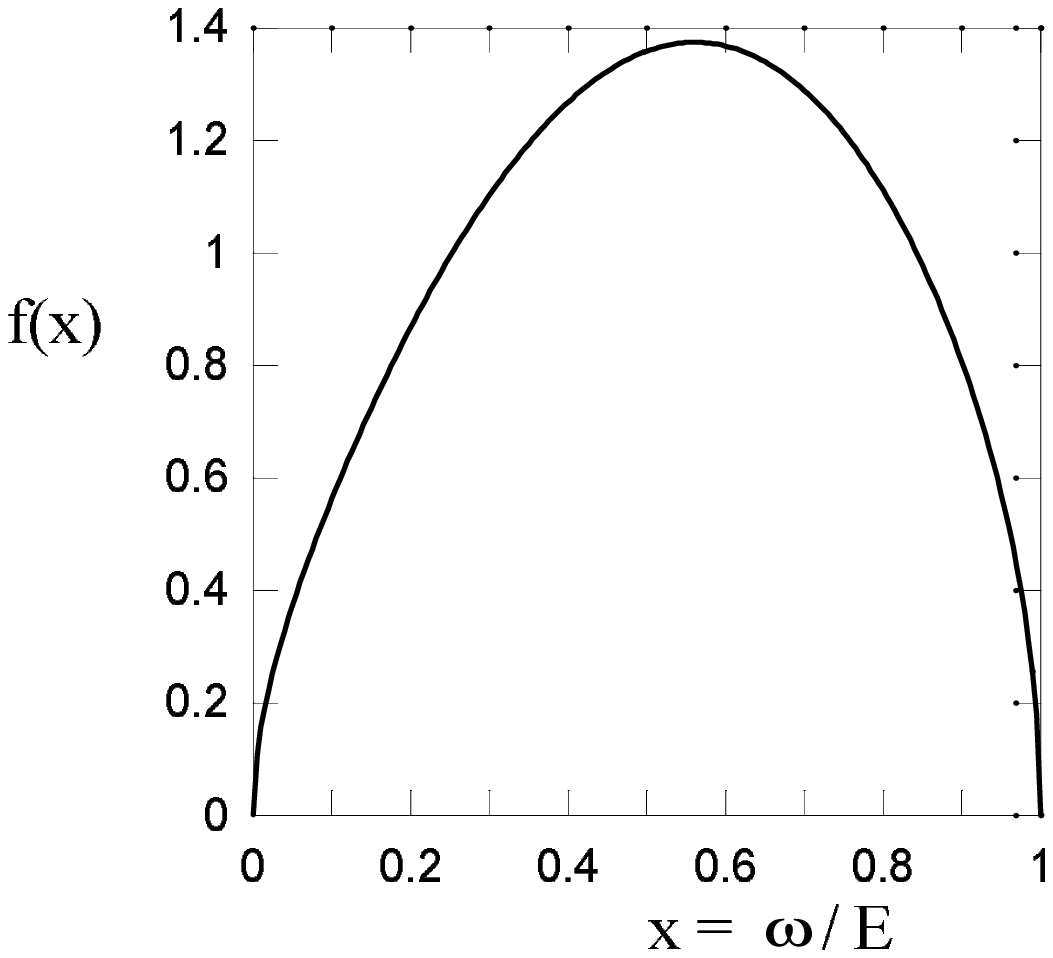}}
\caption[]{
The normalized spectrum of neutrinos $f(x) = (1/ N_\nu)\,
(dN_\nu / dx) $ vs. $x$ --- the fraction of the muon
energy carried by neutrino (see eq.(25)).
}\end{figure}

\begin{figure}[htb]
\epsfxsize=14cm
\centerline{\epsffile{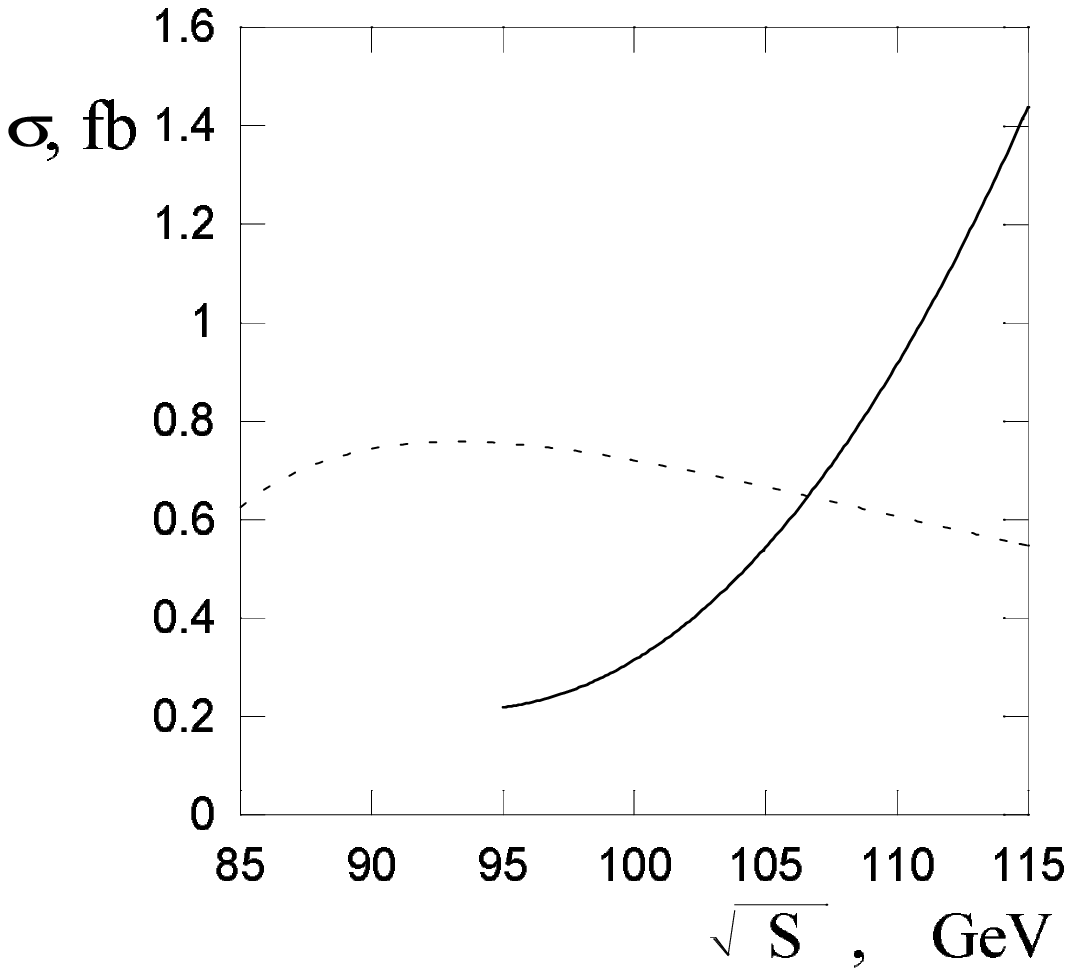}}
\caption[]{
Standard (solid line) and non--standard (dashed line)
contributions to the cross sections (fb) of the reaction $\mu ^-
\mu ^+ \to e \bar \nu _e W^+ $ in dependence on the total cms
energy.  The standard contribution is evaluated with the cut
$-q^2 > m^2$.
}\end{figure}

\end{document}